\documentclass[aps, pra, showpacs, twocolumn, amsfonts, amsmath, amssymb, floatfix,superscriptaddress, groupedaddress]{revtex4}
\usepackage{graphicx}
\usepackage{dcolumn}
\usepackage{bm}
\usepackage{hyperref}
\usepackage{amsmath}

\begin{document}

\title{Behavior of a very large magneto-optical trap}

\author{A.\ Camara$^{1}$, R.\ Kaiser$^{1}$, and G.\ Labeyrie$^{1}$\footnote{To whom correspondence should be addressed.}}
\affiliation{$^{1}$Institut Non Lin\'{e}aire de Nice, UMR 7335 CNRS, 1361 route des Lucioles, 06560 Valbonne, France}

\begin{abstract}
We investigate the scaling behavior of a very large magneto-optical trap (VLMOT) containing up to $1.4 \times 10^{11}$ Rb$^{87}$ atoms. By varying the diameter of the trapping beams, we are able to change the number of trapped atoms by more than 5 orders of magnitude. We then study the scaling laws of the loading and size of the VLMOT, and analyze the shape of the density profile in this regime where the Coulomb-like, light-mediated repulsive interaction between atoms is expected to play an important role.         

\end{abstract}

\maketitle

\section{Introduction}

Since its first realization in 1987~\cite{Raab1987} the magneto optical trap (MOT) has been the working horse of cold atom experiments and continues to be used in a large variety of experiments, such as Bose-Einstein condensates or degenerate Fermi gazes, atomic clocks and sensors and quantum memories. In some of these experiments, increasing the number of trapped atoms is an important advantage. Previous studies have shown that when increasing the number of atoms loaded into a MOT, the peak atomic density tends to saturate and the size of the atomic cloud increases~\cite{Walker1990}. This has been a strong limitation to the straightforward use of the MOT towards Bose-Einstein condensation, requiring novel cooling techniques, often based on conservative trapping potentials combined to evaporation, in order to achieve the quantum degeneracy regime. 
At first it seemed that by increasing the number of atoms in a MOT, a net additional compression force could be obtained due to a shadow effect of the large number of atoms, attenuating the incident laser beams~\cite{Dalibard1988}. However, a more refined model, taking into account the radiation pressure force of the scattered photons showed that the size of a MOT increases with increased atom number~\cite{Walker1990}. This size increase has been confirmed by experiments~\cite{Walker1990} and is due to a modified frequency spectrum of the scattered photons when atoms are driven at large values of the saturation parameter. For a vanishing incident saturation parameter, the shadow effect can at best merely compensate the repulsion force due to the scattered photons, which explains that all experiments up to now have observed an increase in MOT size as the number of atoms is increased. The most commonly used model proposed in ref.~\cite{Walker1990} shows that the repulsion force is analogous to a Coulomb repulsion between particles of same charge, leading to a constant density of particles in a harmonic trap. Further studies \cite{Townsend1995, Kim2004} have shown that in contrast to most common explanations of MOTs, not only the velocity distribution of the trapped atoms but also their spatial distribution might require sub-Doppler mechanisms, such as Sisyphus cooling~\cite{Dalibard1989}. 

With the availability of larger laser power at the relevant wavelengths for cooling and trapping atoms, it is possible to trap more and more atoms in a MOT. It is therefore important to study the MOT scaling laws for large atom numbers, to understand e.g. how the capture velocity and the number of trapped atoms can be maximized. This may allow for adapting designs in new experiments where atom number and trap size are important parameters to be optimized. Another aim would be to obtain experimental signatures in the multiple scattering regime of the MOT that could help improving our understanding of this complex situation and discriminating between various available models. 

We thus report in this paper the results of an experiment where the number of trapped atoms $N$ is varied over a wide range (more than 5 orders of magnitude) in a well-controlled way. The paper is organized as follows. A first section is devoted to the description of the experimental scheme. Details of the experimental procedure are important since it is known to affect the observed scaling laws~\cite{Gattobigio2010}. We then report in section~\ref{loading} our measurement of the scaling law for the number of trapped atoms versus the size of the trapping beams, which is found to increase with an exponent larger than previously reported in the literature. We discuss this result using a simple numerical simulation based on the standard Doppler model for the MOT. We then analyze in Sec.~\ref{size} the scaling law for the size of the MOT as a function of the number of atoms, and compare it to various models. Finally, we discuss in Sec.~\ref{size} the evolution of the shape and ellipticity of the atomic density distribution as the number of atoms is varied. 

\section{Experimental setup}
\label{setup}

\begin{figure}
\begin{center}
\includegraphics[width=1\columnwidth]{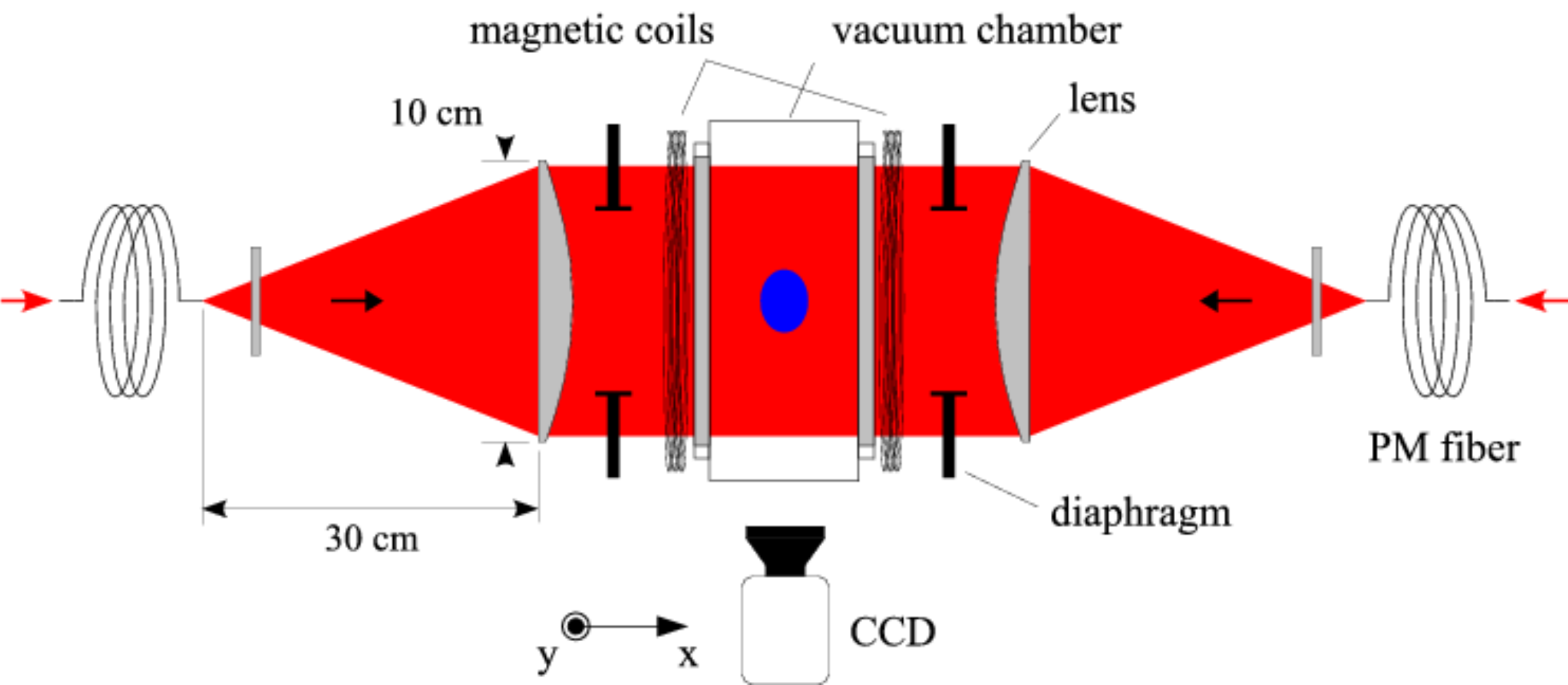}
\caption{(Color on line) Experimental scheme. We show the MOT trapping scheme along one of the three spatial dimensions, corresponding to the axis of the coils generating the magnetic field gradient ($x$). The arrangement is identical for the other two dimensions (except for the magnetic coils). A CCD is used to image the fluorescence of the cold cloud in the ($x, y$) plane (see text for details).}
\label{MOT}
\end{center}
\end{figure}

The six independent VLMOT trapping beams are derived from a single beam using a 1 $\times$ 6 fiber splitter (OZ optics). Both input and output fibers are polarization-maintaining. The input beam is obtained from a weak beam delivered by a DFB laser (2-3 MHz line width), after single-pass amplification through a 2W tapered amplifier. The output fibers tips are placed in the object focal plane of six 10 cm-diameter, 30 cm-focal length lenses to obtain large (waist 2.6 cm) collimated trapping beams (see Fig.~\ref{MOT}). For each dimension of space the corresponding pair of beams is aligned in a counter-propagating fashion. The total trapping power sent to the atoms is 329 mW, corresponding to a peak intensity $I = 5$ mW/cm$^2$ per beam. We trap Rb$^{87}$ using a trapping light detuned by a quantity $\delta_{MOT}$ from the $F = 2 \rightarrow F^{\prime} = 3$ transition. In the present paper, we will use  $\delta_{MOT}$ = -3, -4 or -5 $\Gamma$ (where the natural width $\Gamma = 2 \pi \times 6.06$ MHz). We use these rather large detuning values because they maximize the number of trapped atoms, and also to avoid the dynamical instability that arises at large numbers of atoms and smaller detunings~\cite{Labeyrie2006}. In our setup, the repumping light is produced by another DFB laser tuned close to the $F = 1 \rightarrow F^{\prime} = 2$ transition. The repumper beam is superimposed to the trapping beam (the repumper power representing a few percent of the total) before the injection into the tapered amplifier. Thus, the repumping light is present in each of the 6 VLMOT beams with the same circular polarization as the trapping light, yielding a very symmetrical configuration. The main remaining source of asymmetry in our setup is the slight imbalance between the intensities of the 6 beams, due to the specifications of the fiber-splitter. This imbalance is at most 10$\%$ for two beams in the same counter-propagating pair. A magnetic field gradient of 7.4 G/cm along the axis of the anti-helmoltz coils is applied to spatially trap the atoms.

This experiment aims at measuring scaling laws for the VLMOT as the number $N$ of trapped atoms is varied. We tune $N$ via the diameter of the trapping beams, using six large diaphragms whose aperture $D$ is adjusted (see Fig.~\ref{MOT}). Since the capture range of the MOT depends strongly~\cite{Monroe1990} on $D$, this is an efficient and well-controlled way of varying $N$ without changing the MOT parameters at the location of the trapped atoms.

Fluorescence images are recorded in a plane containing the magnetic gradient coils axis $x$, where the gradient is twice that along the two other axes. We thus have access to the intrinsic anisotropy of the MOT shape, which is studied in section~\ref{shape}. To acquire the fluorescence images, we switch the trapping light detuning to $\delta_{im} = -8 \Gamma$ for a short duration of $230 \mu$s. This is short enough to neglect the displacement of the atoms during the image acquisition ($\approx 30 \mu$m). The large detuning employed to record the fluorescence images has two important consequences: first, the cloud's optical density (OD) at the illuminating light's detuning is then $\ll 1$ (single scattering regime). As a result, the fluorescence intensity distribution closely matches the atomic density distribution (note that this argument is also valid because the effective saturation parameter is only a few $10^{-2}$, which allows one to neglect inelastic scattering and thus the resonant component of the Mollow triplet~\cite{Mollow1969}). As will be discussed in section~\ref{shape}, multiple scattering can strongly distort the recorded fluorescence intensity profiles (see Fig.~\ref{fluo}). Second, because of the large detuning we can safely neglect the Zeeman shift due to the magnetic field gradient which is still on during the measurement (the maximal Zeeman shift across the MOT size is $\approx 1.6 \Gamma$, resulting at most in a $10 \%$ change of the scattering cross-section of the atoms at the edge of the cloud). To improve the signal to noise ratio, we average over 20 successive images. By integrating the fluorescence over the whole images, we get a relative measurement of the number of trapped atoms. A calibration of the absolute number of atoms is performed by measuring the optical density of the cloud along the line of sight of the camera $z$ with a weak probe beam (waist 1.5 mm). For the highest MOT beam diameter $D = 94$ mm and a detuning $\delta_{MOT} = -5 \Gamma$, the on-resonance optical density is 185 and the number of atoms 1.4 $\times 10^{11}$ assuming an equal distribution of the atomic population among Zeeman sub-states.

\begin{figure}
\begin{center}
\includegraphics[angle=-90, width=1\columnwidth]{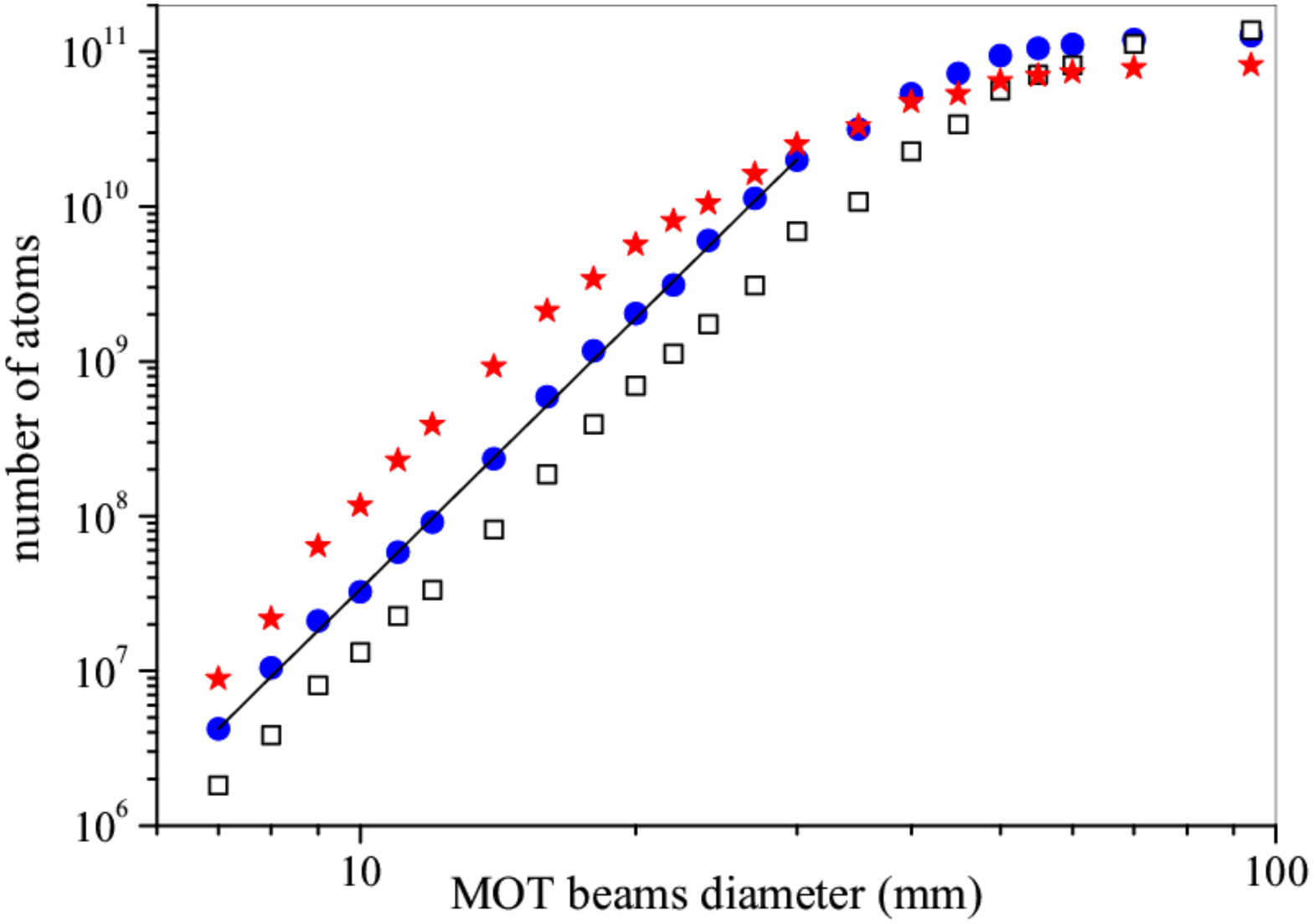}
\caption{(Color on line) Loading the VLMOT (experiment). We plot the number of trapped atoms $N$ as a function of the diameter $D$ of the trapping beams (see text), for three different MOT detunings: $\delta_{MOT} = -3 \Gamma$ (stars); $\delta_{MOT} = -4 \Gamma$ (dots); $\delta_{MOT} = -5 \Gamma$ (squares). We observe a fast increase, followed by a progressive saturation. The line is a fit $N \propto D^{5.82}$ of the data for $\delta_{MOT} = -4 \Gamma$.}
\label{NvsD}
\end{center}
\end{figure}

\section{VLMOT loading}
\label{loading}

Fig.~\ref{NvsD} shows the measured evolution of the number of trapped atoms when the beams diameter is varied (log-log scale). As can be seen, $N$ first increases very strongly with $D$: $N \propto D^{ 5.82}$. This exponent $\alpha = 5.82$ is significantly larger that predicted by the standard model ($\alpha = 4$)~\cite{Monroe1990,Gibble1992}. This Doppler model, based on the balance between loading rate and losses due to collisions with background atoms, leads to:

\begin{equation}
N \propto \frac{D^2}{\sigma} (\frac{v_c}{u})^4
\label{N}
\end{equation}

where $\sigma$ is the collisional cross-section with background atoms, $v_c$ is the velocity capture range of the MOT and $u = \sqrt{2 \frac{k_B T}{m}}$ the most probable velocity in the Maxwell-Boltzmann distribution ($u = 240$ m/s in our case). If one assumes a constant force (i.e. a constant photon scattering rate) acting on an atom inside the MOT volume, one finds~\cite{Gibble1992} $v_c \propto \sqrt{D}$ and from eq.~\ref{N} the scaling $N \propto D^{~4}$ follows (assuming a $v_c$-independent $\sigma$).

\begin{figure}
\begin{center}
\resizebox{1.0\columnwidth}{!}{\includegraphics[angle=-90]{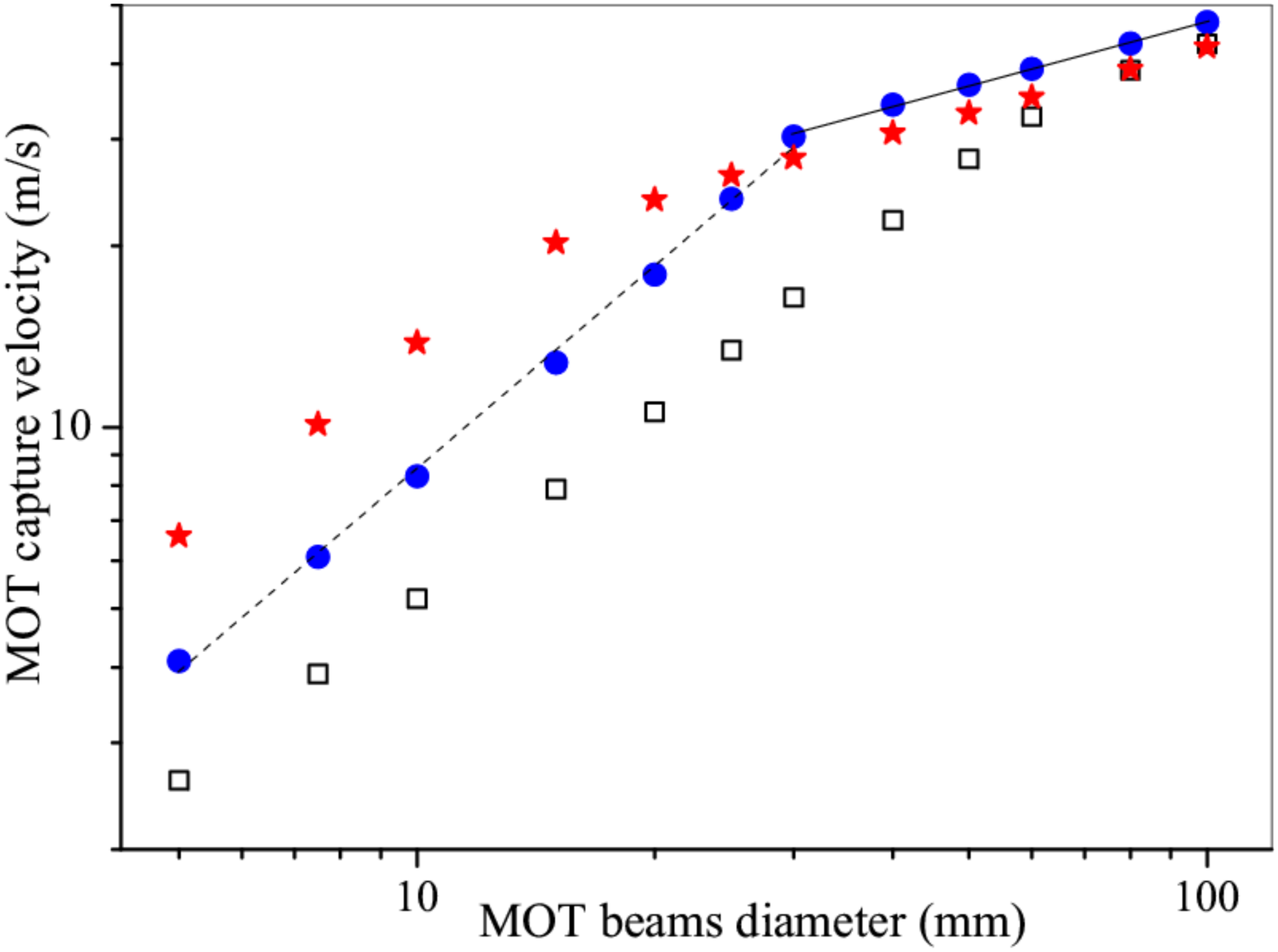}} 
\caption{(Color on line) Capture velocity vs MOT beam diameter (numerics). Using a simple Doppler model, we compute the MOT's capture velocity versus $D$, for the detunings of Fig.~\ref{NvsD}. The lines emphasize the two observed regimes: $v_c \propto D^{~1.1}$ (dotted line) and $v_c \propto D^{~0.36}$ (solid line).}
\label{VcvsD}
\end{center}
\end{figure}

However, the assumption of a constant scattering rate during the trajectory of an atom entering the trapping volume is in general not verified. As the atom moves toward the trap center and is being decelerated, it gradually gets tuned out of resonance with the MOT laser beams and the scattering rate decreases. An accurate estimation of the capture velocity and of its scaling with $D$ thus requires a numerical simulation of the atomic trajectories. We performed such a 3D numerical simulation based on the Doppler model, and found two regimes for the scaling of $v_c$ with $D$ (see Fig.~\ref{VcvsD}): below a certain critical value of $D$, which depends on both $\delta_{MOT}$ and $\nabla B$, $v_c$ is roughly proportional to $D$ (dotted line), while for larger values of $D$ the increase of $v_c$ is slower (solid line). We stress that this cross-over is not due to the finite waist (2.6 cm) of the MOT beams. Instead, it is due to the nonlinear dependency of the MOT force as a function of velocity. For small $D$ the capture velocity is small, and lies in the linear range of the force $k v_c < \left|\delta_{MOT}\right|$. In this regime, increasing $D$ will result in an increase of the capture velocity by roughly the same amount, since the force will increase proportionally to $v_c$. For large $D$ such that $k v_c \approx \left|\delta_{MOT}\right|$, the force is already maximal. Therefore, an increase of $D$ will result in a much smaller increase of $v_c$ than in the linear regime. 

Inserting $v_c~\propto~D$ into Eq.~\ref{N}, we obtain $N \propto D^{~6}$ which is in good agreement with what we measure in Fig.~\ref{NvsD} for $D~<~30$ mm ($N \propto D^{5.82}$). The saturation of the number of trapped atoms at larger $D$ is mainly due to the cross-over seen in Fig.~\ref{VcvsD}, although one expects the Gaussian profile of the MOT beams to enhance this saturation for $D >> w$. Comparing Figs.~\ref{NvsD} and~\ref{VcvsD}, we observe a quite striking qualitative agreement for the behavior of the different detunings. Finally, we note that even higher number of atoms could be loaded in the VLMOT using larger beams and larger detunings, which requires higher laser powers.

\section{VLMOT size scaling}
\label{size}

\begin{figure}
\begin{center}
\resizebox{1.0\columnwidth}{!}{\includegraphics{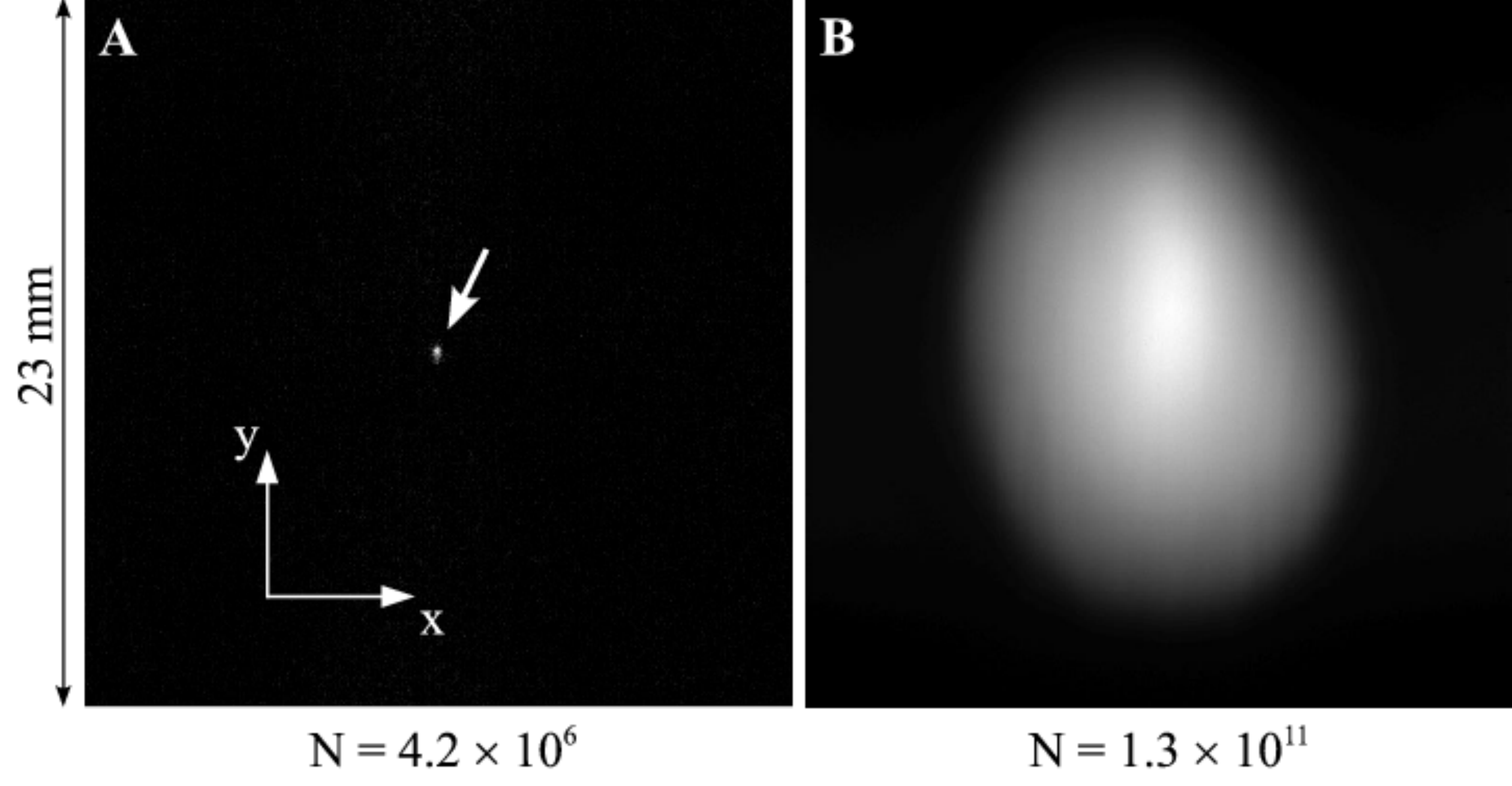}} 
\caption{Variation of VLMOT size with $N$. We show two examples of fluorescence images (see text for details), respectively at low (N = 4.2 $\times 10^6$, \textbf{A}) and large (N = 1.3 $\times 10^{11}$, \textbf{B}) number of trapped atoms. The MOT detuning is $\delta_{MOT} = -4 \Gamma$. The field of view is 23 mm. The axis of the magnetic gradient coils is along x.}
\label{images}
\end{center}
\end{figure}

It is known since the 90s~\cite{Walker1990} that atoms in a MOT are in general not independent, but interact through exchange of photons. The reabsorbtion of scattered photons indeed generates a repulsive inter-atomic force, which tends to expand the cloud. As a result, the size $L$ of the cloud increases with $N$, while it is independent of $N$ in the non-interacting, small-$N$ regime where it is determined only by the MOT parameters and the temperature (hence the name of ''temperature-limited'' regime).

\begin{figure}
\begin{center}
\resizebox{1.05\columnwidth}{!}{\includegraphics[angle=-90]{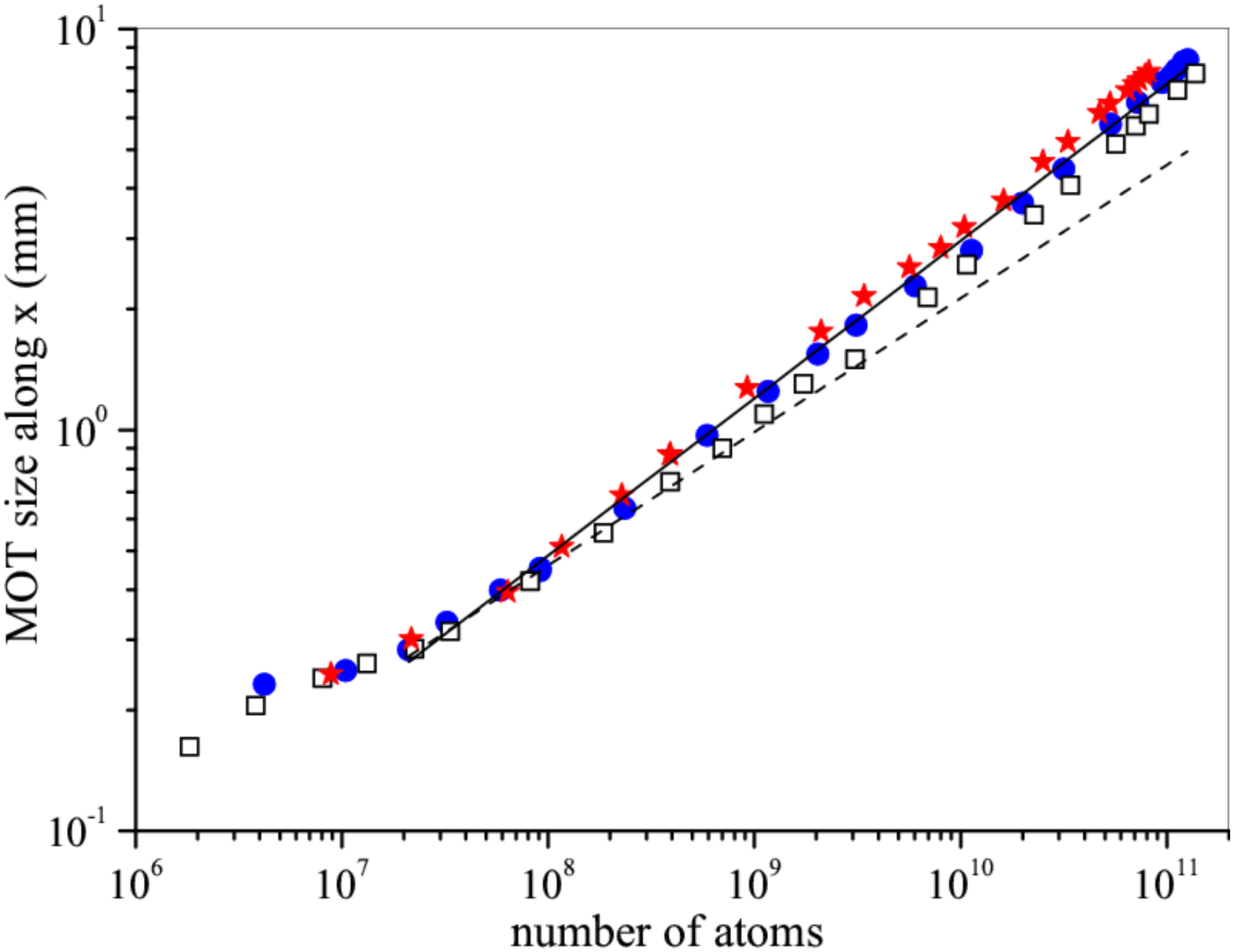}} 
\caption{(Color on line) VLMOT size scaling. We measure the FWHM size of the cloud along the magnetic coils axis $L_x$ as a function of the number of atoms. The three sets of data correspond to different MOT detunings: $\delta_{MOT} = -3 \Gamma$ (stars); $\delta_{MOT} = -4 \Gamma$ (dots); $\delta_{MOT} = -5 \Gamma$ (squares). A fit of the $\delta_{MOT} = -4 \Gamma$ data for $N > 2\times 10^7$ yields $L_x \propto N^{0.39}$ (solid line). The dashed line corresponds to the prediction of the standard model~\cite{Walker1990} $L \propto N^{1/3}$.}
\label{SizevsN}
\end{center}
\end{figure}

\begin{figure}
\begin{center}
\resizebox{1.0\columnwidth}{!}{\includegraphics{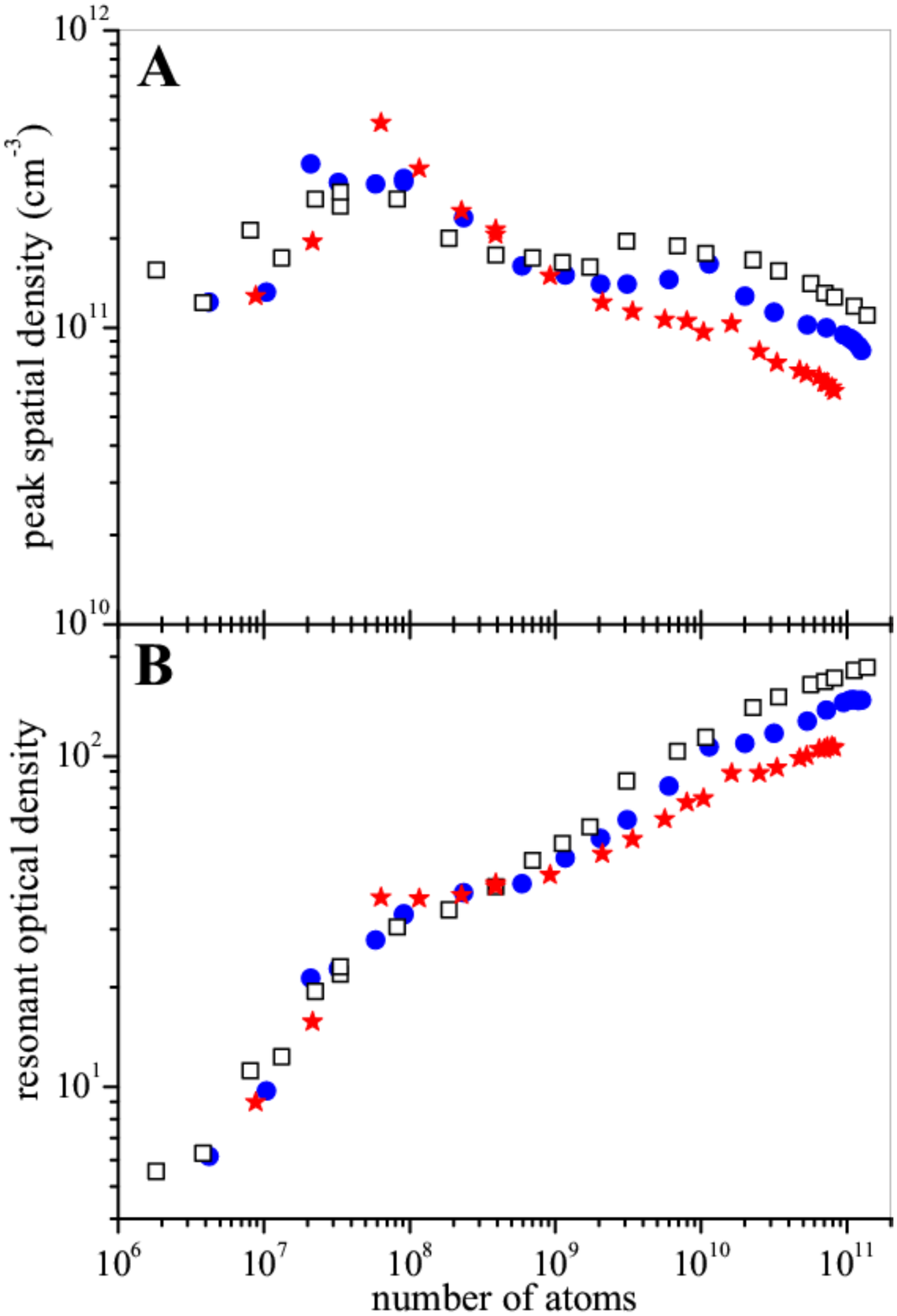}} 
\caption{(Color on line) VLMOT peak density and optical density. We plot in \textbf{A} the peak spatial density and in \textbf{B} the on-resonance optical density of the cloud obtained from the data of Fig.~\ref{SizevsN} (see text).}
\label{n0ODvsN}
\end{center}
\end{figure}

Fig.~\ref{images} illustrates the large variation of MOT size observed in our situation as the number of atoms is tuned. The size varies by a factor $\approx$ 35 while $N$ varies by a factor 31 000. To be more quantitative, we plot on Fig.~\ref{SizevsN} the measured cloud size $L_x$ along the magnetic coils axis, as a function of $N$ and for the three detuning values. This size is determined as the full width at half maximum (FWHM) of a cut of the image, through its center, along x. Since the cloud shape is generally not Gaussian (see next section), we do not integrate the image along y. For $\delta_{MOT} = -4\Gamma$, the size increase for $N > 2\times 10^7$ is well fitted by $L \propto N^{0.394}$. This exponent is observed over a large range of 4 decades. Similar scalings are found for the two other detunings: $L \propto N^{0.388}$ and $L \propto N^{0.411}$ for $\delta_{MOT} = -5 \Gamma$ and $\delta_{MOT} = -3 \Gamma$ respectively. The sizes along the weak confinement axis $y$, not shown in Fig.~\ref{SizevsN}, are larger (see section~\ref{shape}) and exhibit similar exponents: $L \propto N^{0.417}, N^{0.381}$ and $N^{0.35}$ for $\delta_{MOT} = -3, -4$ and $-5 \Gamma$  respectively. For $N < 10^7$, the expansion of the cloud with $N$ seems to slow down. This is indeed expected in the limit of small $N$ where light-induced interactions vanish and the MOT size becomes independent of $N$. However, this regime is expected to occur for much smaller atom numbers: in ref.~\cite{Walker1990}, the temperature-limited regime was observed for $N < 80 000$. It does not seem likely that the observed ''saturation'' of the size around $\approx 200 \mu$m for small $N$ could be due to a poor resolution of our imaging system. The ultimate resolution (limited by the pixel size) is of 23 $\mu$m. Another factor limiting the resolution is the motion of the atoms during the image exposure. The typical displacement corresponding to our temperature is of the order of 30 $\mu$m. The residual effect of multiple scattering is minimized by our choice of a large detuning for the imaging (see fig.~\ref{fluo}), but its exact magnitude remains difficult to estimate. However, its impact is expected to be very small in the small $N$ limit, where the cloud's OD is small (see Fig.~\ref{n0ODvsN}\textbf{B}).

In the standard Doppler model of the MOT~\cite{Walker1990}, the MOT size is determined by the balance between the external trapping force, the inter-atomic repulsion, and a ``shadow'' compressive force due to the attenuation of the MOT beams inside the cloud~\cite{Dalibard1988}. The last two are ``collective'' forces which vanish in the temperature-limited regime. Under the assumptions of ref.~\cite{Walker1990}, which amount to linearizing the trapping and shadow forces and assuming a spatially-independent Coulomb-like interaction force, this balance yields a constant spatial density inside the cloud:
\begin{equation}
n_{max} = \frac{c \kappa}{2 I \sigma_L (\sigma_R - \sigma_L)}
\label{density}
\end{equation}
where $c$ is the speed of light, $\kappa$ is the spring constant characterizing the restoring force for a single-atom MOT, $\sigma_L$ is the absorption cross-section for a laser photon and $\sigma_R$ the cross-section for the absorption of a scattered photon. $\sigma_L$ and $\sigma_R$ are different due to the fact that both the spectral and polarization properties of the scattered light differ from that of the laser light. In this model, increasing $N$ thus result in an expansion of the MOT at constant density: $L \propto N^{~1/3}$. A good agreement with this prediction was reported by the authors of ref.~\cite{Walker1990} for $N < 5\times 10^7$, while they observed a faster increase for larger atom numbers. Possible explanations for this behavior included the effect of the magnetic field gradient and high-order multiple scattering of light inside the cloud. A more involved (numerical) model~\cite{Gattobigio2010} surprisingly led to the same $L \propto N^{~1/3}$ scaling, although the calculated density profiles were no longer homogeneous but displayed a truncated Gaussian shape~\cite{Gattobigio2010}. This model takes into account the nonlinear form of both trapping and shadow forces and the spatial dependence of the interaction force. However the interaction force takes into account only double scattering (a single re-absorption event), as in the standard model. In ref.~\cite{Gattobigio2010}, we have shown that using different techniques to vary the number of atoms (i.e. tuning the intensity or the diameter of a repumping beam) could yield different scaling laws for the MOT size. This still unexplained observation hints at the complexity of the trapping process which is intrinsically multi-level in nature. In the experiment where the diameter of the repumping beam was used as a mean to vary $N$, which is closest in principle from that described in the present paper, a scaling $L \propto N^{0.29}$ was observed which is consistent with the standard model. Our present observation $L \propto N^{0.39}$ is not too far off the $N^{1/3}$ prediction. The complex behavior of the observed MOT shapes, discussed in the next section, may be responsible for this deviation.

Finally, we plot in Fig.~\ref{n0ODvsN}\textbf{A} the peak spatial density of the cloud versus $N$ for the three MOT detunings of Fig.~\ref{SizevsN}. This density is inferred from the measured number of atoms $N$ and sizes $L_x$ and $L_y$, assuming an axially-symmetric MOT $L_z = L_y$ and a Gaussian density distribution. We find densities around a mean value of $2\times 10^{11}$ cm$^{-3}$, which are rather independent of $N$ (typical variation of a factor of 3 over more than 4 orders of magnitude of variation of $N$). The lowest variation of density corresponds to the largest detuning $\delta_{MOT} = -5 \Gamma$, which is due to the fact that the scaling exponent of Fig.\ref{SizevsN} is closest to 1/3. These observations are thus in rough agreement with the constant-density model of~\cite{Walker1990}. Note however that we observe for the cloud's density profiles a different shape (see next section) from that predicted in ~\cite{Walker1990} and measured in ref.~\cite{Sesko1991}. The residual variations observed on the density plot may be attributed to theses changes of cloud shape. Fig.~\ref{n0ODvsN}\textbf{B} shows the on-resonance optical density calculated using the same assumptions. The OD is seen to increase continuously with $N$ with a rough scaling $OD \propto N^{0.3}$ for $\delta_{MOT} = -5\Gamma$ (as determined by a fit over the whole $N$ range) and a maximal value of 185 for our parameters.

\section{VLMOT shape}
\label{shape}

In this section we discuss the evolution of the shape of the cloud as $N$ is varied. Indeed, as emphasized in ref.~\cite{Gattobigio2010}, the density profile of the cloud may be the ultimate signature to discriminate between various models rather than the $L(N)$ scaling. We start by reviewing the existing models in the various MOT regimes, as well as the published observations.

In the limit of small $N$ (temperature-limited regime) where photon re-absorption can be safely neglected, the cloud's density distribution is Gaussian and independent of $N$. For larger atom numbers, when re-absorption sets in, the standard model~\cite{Walker1990} predicts a uniform density profile. This results from the combination of the trapping, ``shadow'' and repulsive forces, with the model of ref.~\cite{Walker1990} assuming a linear spatial dependence of both the first two compression terms and a spatially-independent repulsive force. It has been shown in ref.~\cite{Pohl2006,Gattobigio2010} that including the full spatial dependence of these forces in the Doppler model yields density profiles that are truncated Gaussians. The size $\sigma_{dens}$ of these Gaussians is only determined by MOT parameters, while the truncation radius $R_{tr}$ depends on the number of atoms. In the limit of small $N$ ($R_{tr} \ll \sigma_{MOT}$) one recovers a uniform density profile has predicted by the standard model. On the contrary, in the limit of very large $N$ this spatially-dependent model predicts a Gaussian shape for the density profile.

These predictions rely on the Doppler model of the MOT. However, it was realized very early after the advent of the MOT that sub-Doppler mechanisms play a determinant role in the force near the center of the trap~\cite{Dalibard1989}. This picture, initially developed in the framework of independent atoms, somewhat survives in the regime of multiple scattering albeit with a modified friction and diffusion rate leading to higher temperatures~\cite{Drewsen1994}. When the number of atoms is further increased, the position-dependent profile of the restoring force leads to a ``two-component'' regime for the MOT~\cite{Steane1992,Townsend1995}. There, a central part with a higher density of atoms is subjected to a highly restoring sub-Doppler force, and is surrounded by a halo of lower density where the force is essentially Doppler-like. The radius $R_2$ of the surface separating these two volumes is given by the equality of Zeeman shift and light shift of the ground state~\cite{Townsend1995}:

\begin{equation}
R_2 \approx \frac{\hbar \Omega^2}{\mu_B \nabla B \delta_{MOT}}
\label{R_c}
\end{equation}

If the radius of the cloud is larger than $R_2$, the MOT is in the 2-component regime. For $\nabla B = 20$ G/cm, $\delta_{MOT} = -8 \Gamma$, $I = I_{sat}$ and with the parameters of Cesium, the authors of ref.~\cite{Townsend1995} find that this occurs for $N \approx 10^7$ (with $R_2 \approx 110 \mu$m). Eq.~\ref{R_c} shows that for moderate $N$ where the MOT size is not very large, the two-component regime may be reached for high magnetic gradients and light detunings, and low Rabi frequencies. 

\begin{figure}
\begin{center}
\resizebox{1.0\columnwidth}{!}{\includegraphics[angle=-90]{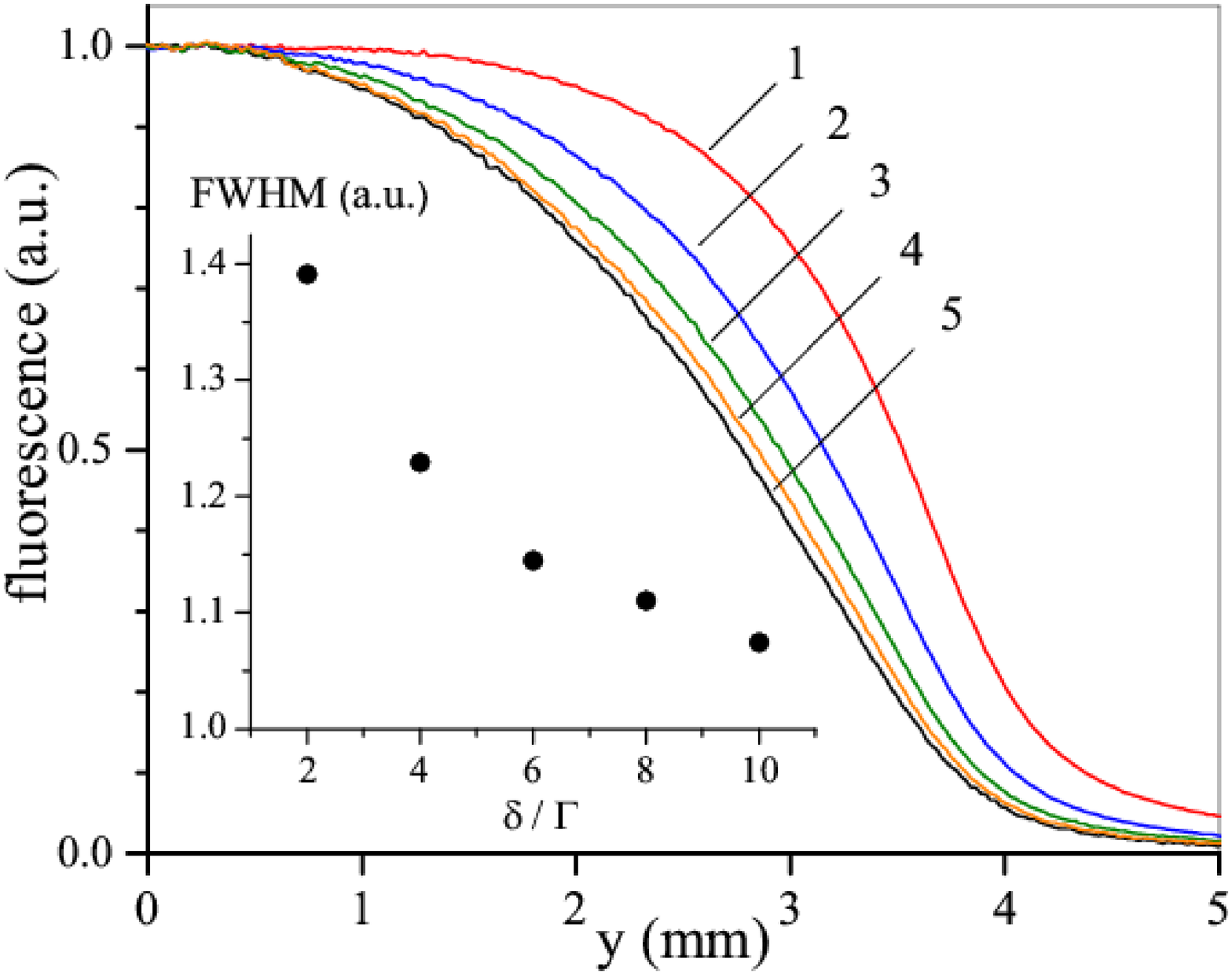}} 
\caption{(Color on line) Impact of multiple scattering during imaging. We compare the fluorescence profiles obtained for the same cloud ($N = 2 \times 10^{10}$) but with different detuning values used for the imaging: $\delta_{im} = -2\Gamma$ (1), $\delta_{im} = -4\Gamma$ (2), $\delta_{im} = -6\Gamma$ (3), $\delta_{im} = -8\Gamma$ (4) and $\delta_{im} = -10\Gamma$ (5). The inset shows the evolution of the measured FWHM with $\delta_{im}$.}
\label{fluo}
\end{center}
\end{figure}

We now turn to the reported measurements of density distributions in a MOT. We first stress that all these were performed by direct fluorescence imaging of the MOT (i.e. using the actual MOT detuning for the imaging), which may cause significant distortions of the profiles as shown in Fig.~\ref{fluo}. Here we compare the profiles obtained for the \textit{same} cloud, but with different values of the detuning $\delta_{im}$ \textit{during the imaging}. The profiles obtained close to resonance are broader with a flatter top than for a detuned illumination, where the profiles become almost independent of $\delta_{im}$ and converge toward the atomic density distribution. The choice of the detuning also significantly affects the measured width, as illustrated in the insert. We thus conclude that when the shape measurement is performed by direct imaging of the MOT fluorescence (and not with a detuned excitation as done here), one should be cautious with the interpretation of the recorded profiles as long as the OD at the MOT detuning is not $\ll 1$ (in Fig.~\ref{fluo}, it is 0.4 for $\delta_{im} = -8 \Gamma$).

A general feature of most reported cloud shape measurements (including ours) is that the density distribution is integrated along the line of sight of the detection device. For an axially-symmetric MOT it is in principle possible to reconstruct the 3D density distribution using an inverse Abel transformation~\cite{Overstreet2005}, but since its implementation necessitates low noise and highly symmetrical MOT shapes it is in general unpractical.

The authors of ref.~\cite{Sesko1991} reported Gaussian profiles corresponding to the temperature-limited regime for $N < 8\times 10^4$. For larger $N$, the standard model predicts a constant density which integrated once yields a profile $f(x) \propto \sqrt{R^2 - x^2}$ where $R$ is the radius of the uniform sphere of atoms. Such rather flat profiles were also observed in ref.~\cite{Sesko1991}, but not in a subsequent detailed study~\cite{Townsend1995} where Gaussian profiles were observed instead. The ``constant density'' signature of the multiple-scattering regime was then observed on the peak density, similarly to what we show in Fig.~\ref{n0ODvsN}\textbf{A}. Deviations from a Gaussian were also reported in ref.~\cite{Grego1996}, and well fitted to the functional dependence introduced by the authors of ref.~\cite{Hoffmann1994} to account for multiple scattering and finite temperature. The difference between all these experimental findings is not elucidated, but we note that flat-top profiles can also be due to multiple scattering of the illuminating light even if the density distribution is Gaussian, as discussed before. The two-component regime was observed in ref.~\cite{Petrich1994,Townsend1995}, and the sub-Doppler component was nicely separated from the Doppler halo in ref.~\cite{Kim2004}.

\begin{figure}
\begin{center}
\resizebox{1.05\columnwidth}{!}{\includegraphics{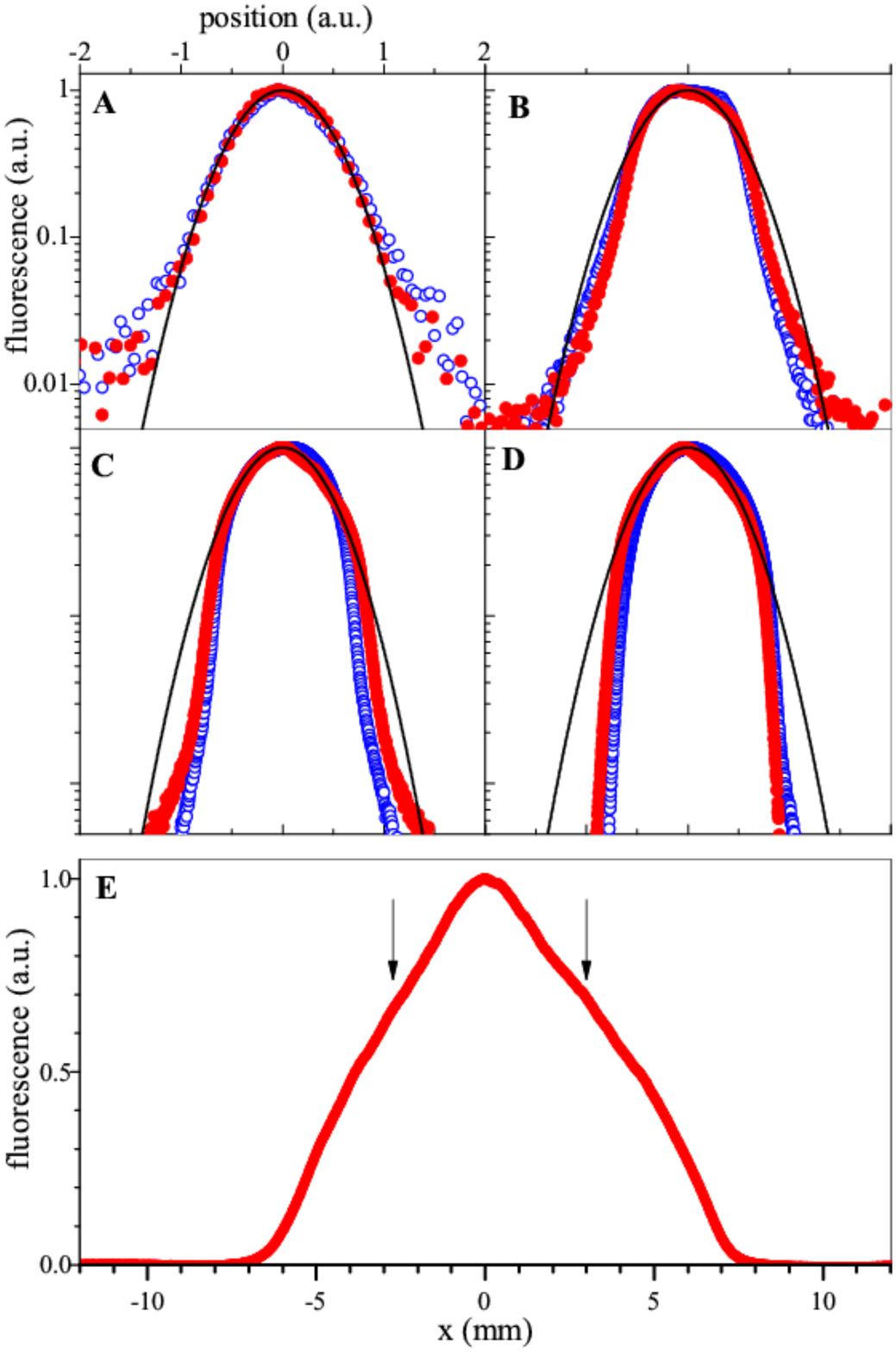}} 
\caption{Fluorescence profiles of the cloud. We plot the fluorescence profiles along $x$ (dots) and $y$ (circles) for four different atom numbers: (\textbf{A}) $N = 3.2 \times 10^7$; (\textbf{B}) $N = 1.2 \times 10^9$; (\textbf{C}) $N = 2 \times 10^{10}$; (\textbf{D}) $N = 1.3 \times 10^{11}$. The lines correspond to a Gaussian shape. The data of panels (\textbf{A}) to (\textbf{D}) are all scaled in the same way: all profiles are vertically normalized to a maximum value = 1 (note the log scale), while the horizontal scaling is different for all four plots such that the profile's FWHM is equal to 1. Panel (\textbf{E}) shows the data of (\textbf{D}) along $x$, in linear scale. }
\label{profiles}
\end{center}
\end{figure}

We now discuss our observations. We show in Fig.~\ref{profiles} some fluorescence profiles recorded for different atom numbers (panels \textbf{A} to \textbf{D}) at $\delta_{MOT} = -4 \Gamma$. These profiles are cuts of the 2D fluorescence images such as shown in Fig.~\ref{images} along the two axes $x$ and $y$. The symbols correspond to the data, the lines to Gaussian profiles. The vertical and horizontal scales are normalized to ease the comparison. The horizontal scaling is different in panels \textbf{A} to \textbf{D}, and is chosen such that the FWHM of the profiles is equal to 1. The vertical scale is logarithmic to allow for a better observation of the wings, and the scaling is such that the peak value of the profiles is 1. For $N$ below typically $10^8$ atoms, we obtain profiles quite close to Gaussians (Fig.~\ref{profiles}\textbf{A}). When $N$ is increased to roughly $10^9$ (Fig.~\ref{profiles}\textbf{B}), the profiles deviate from a Gaussian and get quite close to the flat-top shapes of ref.~\cite{Hoffmann1994, Grego1996}. This is accompanied by a steepening of the wings of the profiles, which is a prediction of all models including multiple scattering~\cite{Walker1990,Hoffmann1994,Gattobigio2010}. When $N$ is increased even further, the profile along $y$ gradually rounds off, while the profile along $x$ develops for $N > 10^{10}$ a central feature with enhanced density (Fig.~\ref{profiles}\textbf{C} and \textbf{D}). This last behavior is best seen in Fig.~\ref{profiles}\textbf{E} where we plot the data of \textbf{D} along $x$ in linear scale (the arrows point at the inflexion points in the profile). We stress that this general behavior is, apart from minor details in the shapes, robust against modifications of the MOT alignment such as e.g. the beam intensity imbalance. We find that all profiles along $x$ for $N$ between $10^{10}$ and $10^{11}$ are consistent with a double-component distribution, including a narrower part near the MOT center. We believe that for this range of atom numbers, our MOT operates in the two-component regime. Indeed, eq.~\ref{R_c} yields in our case $R_2 \approx 1$ mm (with $\nabla B = 7.4$ G/cm, $\delta_{MOT} = -4 \Gamma$ and $\Omega^2 / \Gamma^2 = 0.7$). This corresponds to $N \approx 4 \times 10^9$, which is in rough agreement with the appearance of the central feature. This is also in rough agreement with an extrapolation of the MOT ``phase diagram'' computed in ref.~\cite{Townsend1995}. A measurement of the velocity distribution of the atoms, not performed in this work, could possibly corroborate this hypothesis. We also find that for a fixed $N$, the deviation from a Gaussian profile is larger when the detuning is smaller (MOT operating closer to resonance), which is to be expected for multiple scattering effects.

\begin{figure}
\begin{center}
\resizebox{1.0\columnwidth}{!}{\includegraphics[angle=-90]{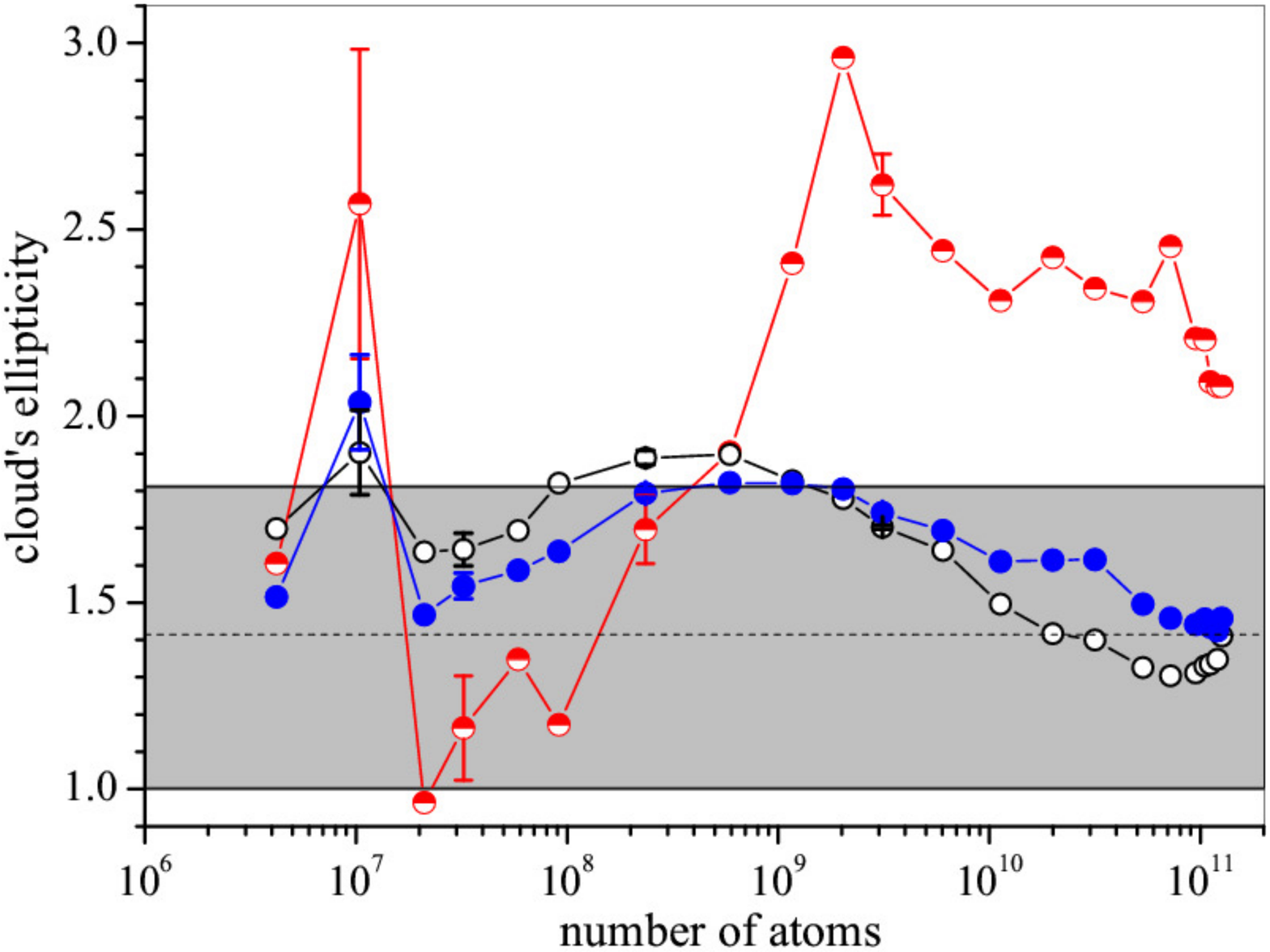}} 
\caption{(Color on line) Ellipticity of the cloud. We plot the ellipticity $\epsilon$ of the MOT measured versus the number of atoms $N$. $\epsilon$ is measured at $90 \%$(half-filled circles), $50 \%$(dots) and $10 \%$(circles) of the peak value of the fluorescence images. The shaded area corresponds to the limits obtained from the model of ref.~\cite{Romain2014}, while the dashed line $\epsilon = \sqrt{2}$ is the expected ellipticity for a MOT in the temperature-limited regime.}
\label{AR}
\end{center}
\end{figure}

The double-component behavior does not show up clearly in the profiles along $y$. Indeed, one expects from eq.~\ref{R_c} that the radius of the sub-Doppler central feature is inversely proportional to the magnetic field gradient. We thus expect for the central feature an ellipticity $\epsilon = L_y / L_x = 2$. However, the ellipticity of the Doppler component which makes up most of the cloud's size is only of the order of 1.5 as can be seen on Fig.~\ref{AR}. The widths of the sub-Doppler and Doppler components are thus more similar along $y$, rendering their differentiation difficult. Fig.~\ref{AR} shows the cloud's ellipticity measured versus $N$, at different proportions of the peak value in the fluorescence images: $90 \%$ (stars), $50 \%$ (dots) and $10 \%$ (squares). It can be seen that the ellipticities measured at $10$ and $50 \%$ of the maximum are following a quite parallel evolution when $N$ is varied, with values around 1.5 for $N > 10^{10}$. On the contrary, the ellipticity measured at $90 \%$ of the maximum (i.e. near the center of the cloud) show a steep increase for $N > 10^9$ and reaches higher values at large $N$ (average of $\epsilon =2.2$ for $N > 10^{10}$). This behavior is consistent with the appearance of a two-component distribution for high $N$ values.

The shaded area on Fig.~\ref{AR} corresponds to the possible values of $\epsilon$ according to the model of ref.~\cite{Romain2014}. The authors of this recent work proposed the measurement of the ellipticity as a mean to determine experimentally the cross-section ratio $\frac{\sigma_R}{\sigma_L}$ (see eq.~\ref{density}), an interesting quantity difficult to compute in a realistic MOT situation. Their model rely on the standard approach of ref.~\cite{Walker1990}, using the same hypothesis (small OD, double scattering only, and spatially-independent cross-sections $\sigma_L$ and $\sigma_R$). It predicts a variation of $\epsilon$ with the MOT parameters (intensity and detuning), but not with $N$ as it is observed in the present work (Fig.~\ref{AR}). This is not surprising, however, since we expect these assumtions to break down at large $N$ values. Furthermore, our complicated MOT shape behavior is clearly not accounted for by this model.

\begin{figure}
\begin{center}
\resizebox{1.05\columnwidth}{!}{\includegraphics[angle=-90]{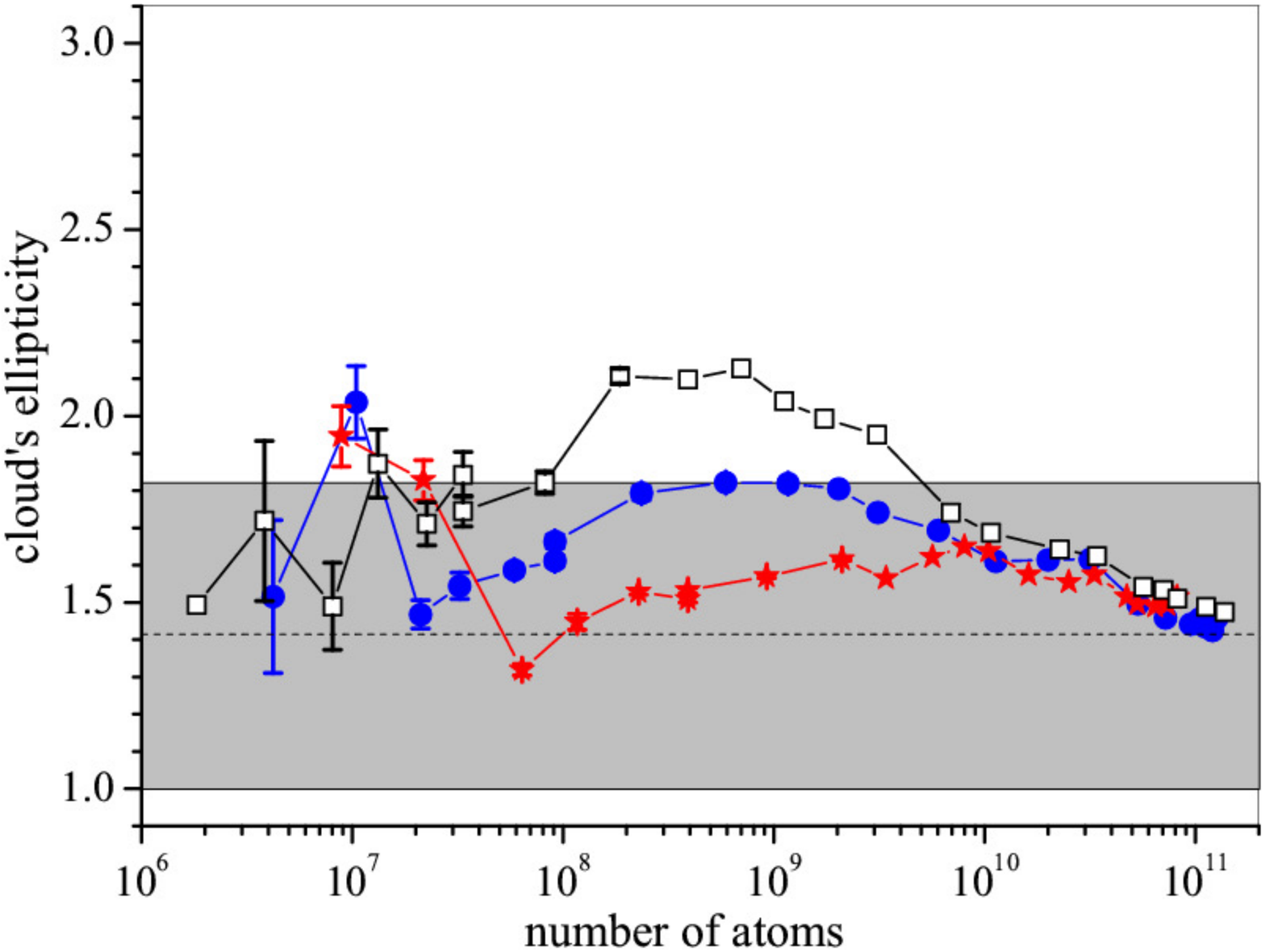}} 
\caption{(Color on line) Ellipticity of the cloud versus MOT detuning. We plot here the ellipticity measured at $50 \%$ of the peak value of the fluorescence images, for different values of $\delta_{MOT}$: $\delta_{MOT} = -3 \Gamma$ (stars); $\delta_{MOT} = -4 \Gamma$ (dots); $\delta_{MOT} = -5 \Gamma$ (squares). The shaded area corresponds to the limits obtained from the model of ref.~\cite{Romain2014}, while the dashed line $\epsilon = \sqrt{2}$ is the expected ellipticity for a MOT in the temperature-limited regime.}
\label{ARvsDelta}
\end{center}
\end{figure}

Fig.~\ref{ARvsDelta} shows how $\epsilon$ (measured at $50\%$ of the peak fluorescence) depends on $\delta_{MOT}$. We observe globally that for intermediate atom number $10^8 < N < 7 \times 10^9, \epsilon$ increases with $\left|\delta_{MOT}\right|$. In the framework of ref.~\cite{Romain2014} this would correspond to a strong increase of $\frac{\sigma_R}{\sigma_L}$ (note however that for $\delta_{MOT} = -5 \Gamma$ our measured $\epsilon$ largely exceeds the theoretical limit of 1.81). Interestingly, for $N > 7 \times 10^9$ all curves collapse together and seem to converge towards the temperature-dependent limit (dashed line). This corresponds roughly to the situation where the cloud's optical density at $\delta_{MOT}$ becomes larger than 1. In this regime which is clearly beyond the reach of the standard model of ref.~\cite{Walker1990}, the trapping laser beams are strongly attenuated inside the cloud. A more refined model needs to be developed to understand how shadow effect and multiple scattering concur to yield the observed behavior. 

\section{Conclusion}
In this paper we have presented our observations on the behavior of a very large magneto-optical trap containing up to $1.4 \times 10^{11}$ atoms. To our knowledge, this is the largest number of atoms in a MOT reported in the literature. The number of trapped atoms and the cloud's size and shape are studied as a function of the diameter $D$ of the MOT's lasers beams. Using this technique, the atom number can be varied by 5 orders of magnitude. We observe an increase of $N$ with $D$ much faster than previously reported, a feature well-reproduced by simulations of the MOT's capture velocity based on a simple Doppler model. We find a scaling of the cloud size versus $N$ roughly consistent with the standard model of a MOT in the multiple scattering regime, even up to such large numbers of atoms. A careful measurement of the cloud shape yields Gaussian profiles up to $10^8$ atoms, and then strong deviations for larger $N$. For $N > 10^{10}$, our observations are consistent with the two-component regime for the MOT, in agreement with the predictions of ref.~\cite{Townsend1995}. Such large MOTs where strong multiple scattering effects constitute interesting tools to search for analogies with e.g. plasma physics, hydrodynamics, or stellar physics~\cite{Labeyrie2006}. They can also be used to produce large (centimeter-scale) cold clouds with a high optical density, well-suited to perform original experiments in e.g. nonlinear optics~\cite{Labeyrie2011} and self-organization~\cite{Labeyrie2014}.

\acknowledgments{This work was supported by CNRS and Universit\'e de Nice-Sophia Antipolis. We also acknowledge financial support from R\'egion PACA, F\'ed\'eration Wolfgang Doeblin.}

\end{document}